\documentclass[fleqn,usenatbib]{mnras}

\usepackage{newtxtext,newtxmath}

\usepackage[T1]{fontenc}
\usepackage{ae,aecompl}

\usepackage{graphicx}	
\usepackage{amsmath}	
\usepackage{amssymb}	
\usepackage{float}
\usepackage{fancyhdr}
\usepackage{hyperref}
\usepackage{longtable}
\usepackage{subfig}
\usepackage{graphicx}
\usepackage[usenames,dvipsnames]{xcolor}

\usepackage{soul}

\usepackage{epstopdf}

\newcommand{\Teff}{\mbox{$T_{\mathrm{eff}}$}}

\newcommand{\Lines}[3]{\Ion{#1}{#2}\,#3\,\AA}
\newcommand{\Ion}[2]{#1\,{\sc #2}}
\newcommand{\La}{\mbox{${\mathrm{Ly\alpha}}$}}
\newcommand{\Ha}{\mbox{${\mathrm{H\alpha}}$}}

\newcommand{\Msun}{\mbox{$\mathrm{M}_{\odot}$}}

\newcommand{\HHe}{\mbox{$\log(N_\mathrm{H}/N_\mathrm{He})$}}

\title[Broadening of \La\ by neutral helium]{Broadening of \La\ by neutral helium in DBA white dwarfs}
 
\author[B.T. G\"ansicke et al.]{
Boris T. G\"ansicke$^{1}$\thanks{E-mail: Boris.Gaensicke@warwick.ac.uk},
Detlev Koester$^{2}$,
Jay Farihi$^{3}$,
Odette Toloza$^{1}$\\
$^{1}$ Department of Physics, University of Warwick, Coventry CV4 7AL, UK\\
$^{2}$ Institut f\"ur Theoretische Physik und Astrophysik, University of Kiel, 24098 Kiel, Germany\\
$^{3}$ Department of Physics and Astronomy, University College London, London WC1E 6BT, UK
}

\date{Accepted XXX. Received YYY; in original form ZZZ}

\pubyear{2018}

\begin{document}
\label{firstpage}
\pagerange{\pageref{firstpage}--\pageref{lastpage}}
\maketitle

\begin{abstract}
Traces of photospheric hydrogen are detected in at least half of all white dwarfs with helium-dominated atmospheres through the presence of \Ha\ in high-quality optical spectroscopy. Previous studies have noted significant discrepancies between the hydrogen abundances derived from \Ha\ and \La\ for a number of stars where ultraviolet spectroscopy is also available. We demonstrate that this discrepancy is caused by inadequate treatment of the broadening of \La\ by neutral helium. When fitting \textit{Hubble Space Telescope} COS spectroscopy of 17 DB white dwarfs using our new line profile calculations, we find good agreement between \HHe\ measured from \La\ and \Ha. Larger values of \HHe\ based on \La\ are still found for three stars, which are among the most distant in our sample, and we show that a small amount of interstellar absorption from neutral hydrogen can account for this discrepancy. 

\end{abstract}




\section{Introduction}
White dwarfs are the progeny of stars with initial masses $\la 8-10\,\Msun$ \citep[e.g.][]{garcia-berroetal97-1, smarttetal09-1, dohertyetal15-1}. The high surface gravities of these stars result in the chemical stratification of their thin non-degenerate envelopes \citep{schatzman48-1}. Most white dwarfs have a sufficiently thick layer of hydrogen to result in optical and ultraviolet spectra dominated by Balmer and Lyman lines, respectively, and are spectroscopically classified as DA stars \citep{sionetal03-1}. However, a small fraction of white dwarfs have helium-dominated atmospheres, probably a result of a late thermal pulse removing the residual hydrogen \citep{ibenetal83-1, althausetal15-1}. For effective temperatures $\Teff\ga10\,000$\,K, the spectra of these stars contain absorption lines of neutral helium (spectral type DB), at lower temperatures their spectra are featureless continua (DC white dwarfs). The relative fraction of white dwarfs with helium-dominated atmospheres is a function of effective  temperature, and hence of cooling age. The changes in this fraction at the hotter end of the cooling sequence are generally understood by mixing of thin hydrogen layers in a increasingly deep convection zone. Variations of the ratio of helium vs. hydrogen dominated atmospheres near the cool end of the cooling sequence are less well understood \citep{bergeronetal97-1, chenetal12-1, giammicheleetal12-1}.

\citet{liebertetal79-1} showed that the dichotomy between helium vs. hydrogen dominated atmospheres is not clean, identifying WD\,1425+540 (G200-39) as a relatively cool ($\simeq15\,000$\,K) white dwarf exhibiting Balmer and helium absorption lines of roughly equal strength with an estimated $\HHe\simeq-3.6$. Shortly later, \citet{liebertetal84-1} discovered a hotter ($\simeq30\,000$\,K) example of a white dwarf with both Balmer and helium lines, showing that trace hydrogen in helium-atmosphere white dwarfs can be present across a wide range of effective temperatures. While these stars are classified as DAB or DBA white dwarfs depending on the relative strength of the Balmer and helium lines, it is important to note that the spectra of some helium-atmosphere white dwarfs, in particular cooler ones, are dominated by Balmer lines, e.g. GD\,362 \citep{gianninasetal04-1, kawka+vennes05-1}, GD\,16 \citep{koesteretal05-1}, and GD\,17 \citep{gentile-fusilloetal17-1}. At higher temperatures, trace helium in hydrogen-dominated atmospheres becomes detectable via the presence of \ion{He}{I} lines \citep{manseauetal16-1}. Whereas a small number of these mixed atmosphere stars turned out to be unresolved DA plus DB binaries (e.g. WD\,1115+166, \citealt{bergeron+liebert02-1}), surveys with better instrumentation now  demonstrate that a large fraction of helium atmosphere white dwarfs contain varying amounts of trace hydrogen \citep{vossetal07-1, bergeronetal11-1,koester+kepler15-1, rollandetal18-1}. 

\begin{table}
  \centering
\caption{Log of the \textit{HST}/COS observations. }
\begin{tabular}{cccc}
\hline
\hline
WD & Date & Exposure & Program \\
& & time\,[s] & ID \\
\hline
0100--068 & 2011-10-25 & 800  & 12474 \\
0110--565 & 2017-03-10 & 5734 & 14597 \\
0125--236 & 2017-01-23 & 5266 & 14597 \\
0435+410  & 2012-01-28 & 1600 & 12474 \\
0437+138  & 2017-09-07 & 5262 & 14597 \\
1107+265  & 2017-05-26 & 8116 & 14597 \\
1349--230 & 2016-08-27 & 8077 & 14597 \\
1352+004  & 2017-07-15 & 8041 & 14597 \\
1425+540  & 2014-12-08 & 8358 & 13453 \\
1557+192  & 2012-01-29 & 1460 & 12474 \\
1644+198  & 2017-07-14 & 8088 & 14597 \\
1822+410  & 2012-01-11 & 1400 & 12474 \\
1940+374  & 2011-12-29 & 1460 & 12474 \\
2144--079 & 2011-11-13 & 1600 & 12474 \\
2229+139  & 2017-10-24 & 13905 & 14597\\
2354+159  & 2017-07-23 & 2187 & 14597 \\
\hline
\end{tabular}
\label{tab:obs}
\end{table}

The origin of this trace hydrogen has been explored extensively \citep{beauchampetal96-1, vossetal07-1, bergeronetal11-1, koester+kepler15-1, rollandetal18-1}, however, the two favoured explanations, a residual thin hydrogen layer or accretion from the interstellar medium face problems explaining the apparent change in \HHe\ as a function of cooling age. Episodic accretion of water-bearing planetesimals has recently been proposed as an alternative source of hydrogen \citep{farihietal10-2, farihietal13-2, raddietal15-1, gentile-fusilloetal17-1}, and it is remarkable to note that \citet{xuetal17-1} showed that the prototypical DBA WD\,1425+540 is accreting volatile-rich planetary material, with a composition similar to solar system comets. 

In the vast majority of DBA and DAB stars, the hydrogen content has been derived from the analysis of optical spectroscopy, in particular the strength of the \Ha\ line \citep{vossetal07-1, bergeronetal11-1, koester+kepler15-1, rollandetal18-1}. Ultraviolet spectroscopy is available for a number of DB stars, allowing an independent assessment of \HHe\ from the \La\ absorption line, and \citet{jura+xu12-1} noticed significant discrepancies between the values of \HHe\ derived from \Ha\ and \La. The most extreme published example of differing \HHe\ derived from optical and ultraviolet spectroscopy is WD\,1425+540 \citep{genest-beaulieu+bergeron17-1}. However, so far no systematic comparison between the values derived from \Ha\ and \La\ has yet been attempted.

Here we present a homogeneous analysis of \textit{Hubble Space Telescope} (\textit{HST}) ultraviolet spectroscopy of 17 DB white dwarfs, making use of new calculations of the \La\ profile broadened by neutral helium. 

\section{Observations}
The far-ultraviolet spectroscopy was obtained as part of three separate \textit{HST} programs (\#12474, \#13453, and \#14597) between October 2011 and October 2017 (Table\,\ref{tab:obs}). All observations were carried out using the G130M grating with a central wavelength of 1291\,\AA, except for WD\,0125--236, which was observed with a central wavelength of 1327\,\AA. The 1291\,\AA\ observations covered the wavelength range $1130-1435$\,\AA, with a gap at $1278-1288$\,\AA\ due to the space between the two detector segments. For the 1327\,\AA\ observations, the wavelengths covered were $1170-$1470\,\AA, with a gap $1318-1328$\,\AA. We used all four FP-POS positions to mitigate the fixed pattern noise that is affecting the COS far-ultraviolet detector for the observations obtained in programs \#13453 and \#14597. The observations obtained in program \#12474 were part of a snapshot survey of white dwarfs, and the short exposure times allowed only two FP-POS positions to be used \citep{gaensickeetal12-1}. To mitigate airglow emission, we replaced the region around \Lines{O}{I}{1302,04} with the spectrum extracted from the COS data obtained on the night side of the Earth.

\begin{figure}
\centerline{\includegraphics[width=\columnwidth]{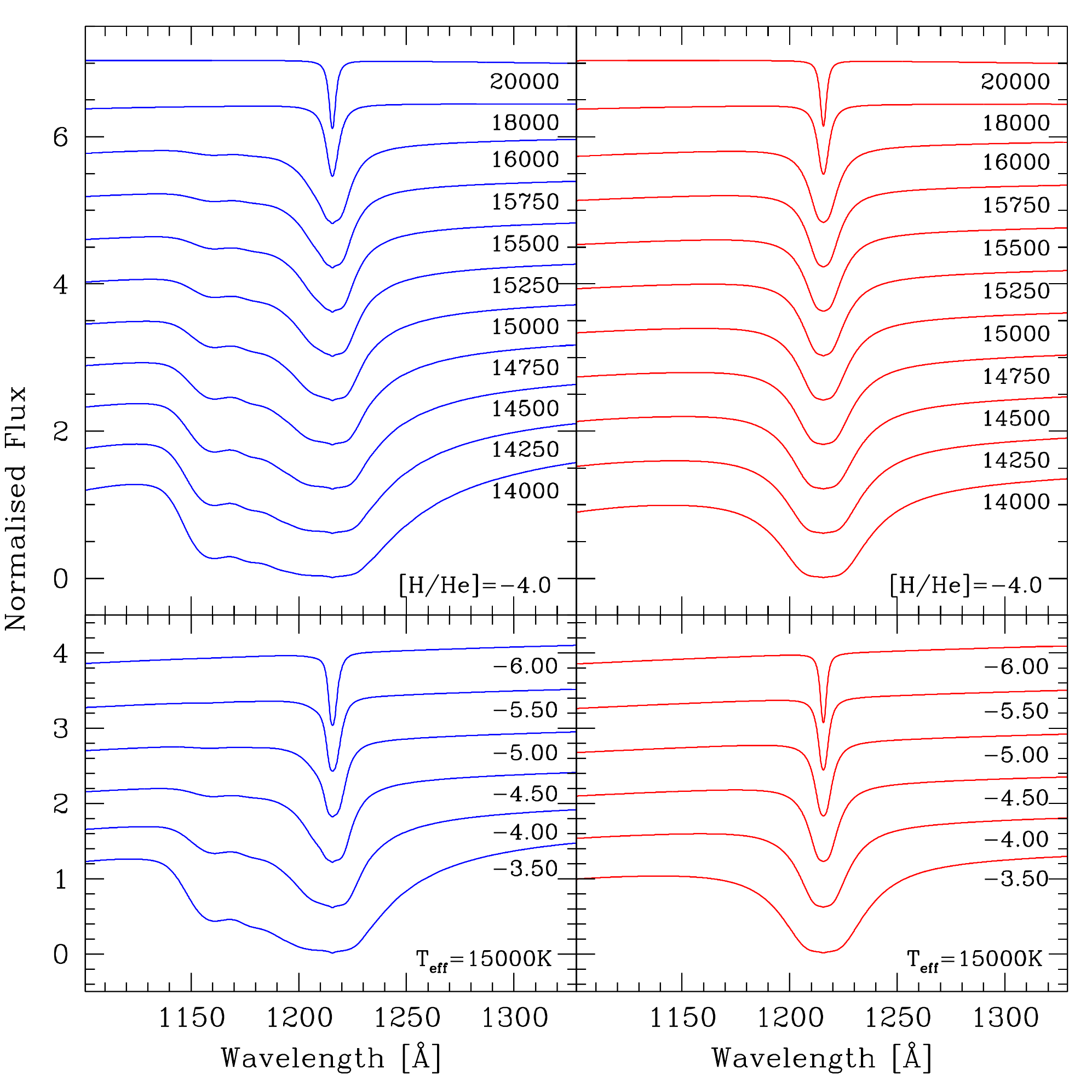}}
\caption{\label{fig:profiles} \La\ line profiles computed for a range of effective temperatures and values of \HHe, using van der Waals broadening by neutral hydrogen and helium (right, red lines) and using our improved unified line broadening theory (left, blue lines). All models were computed with $\log g=8$ fixed, and the normalised spectra were offset vertically by multiples of 0.6 units.}
\end{figure}

\begin{figure}
\centerline{\includegraphics[width=\columnwidth]{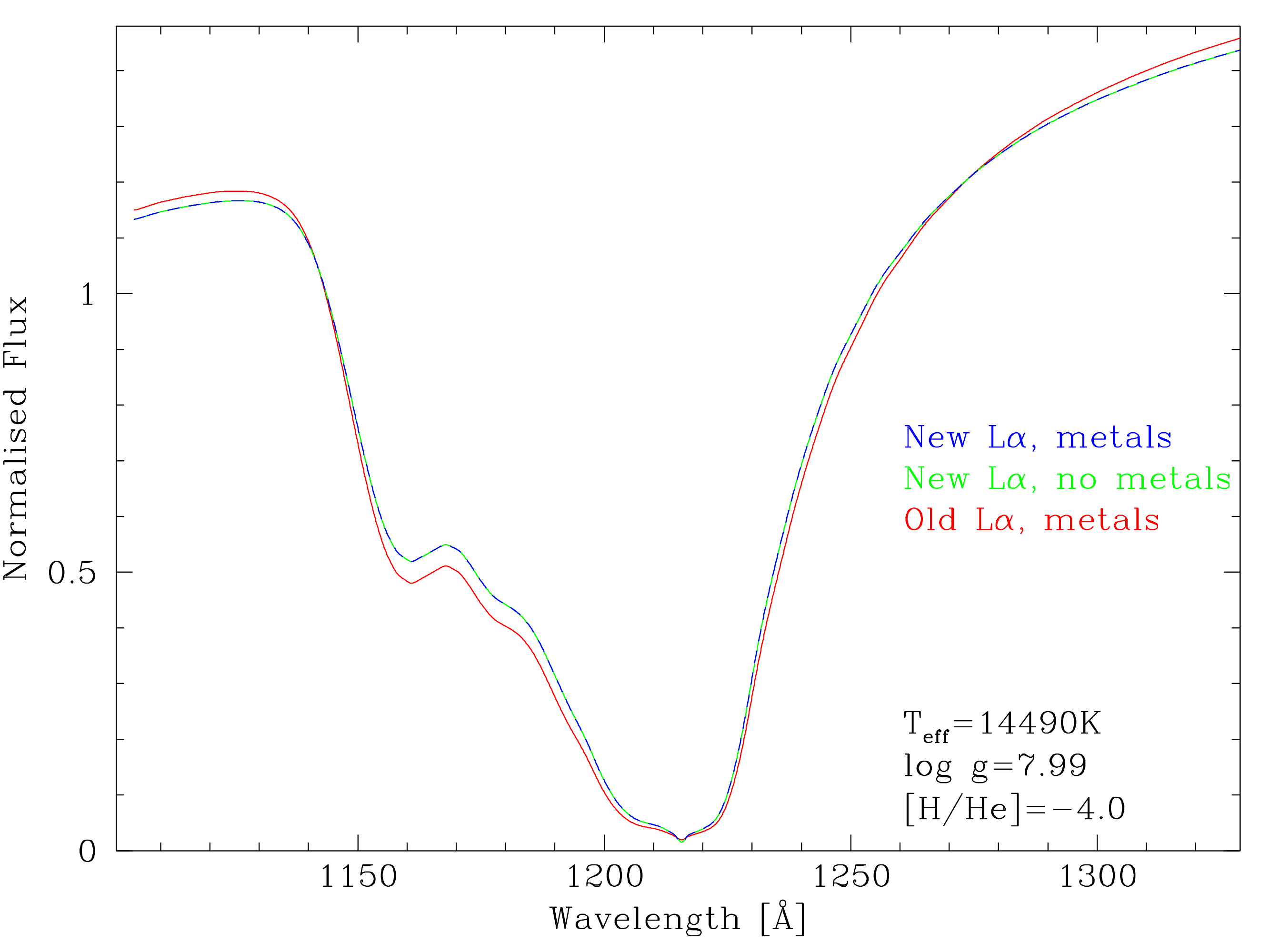}}
\caption{\label{fig:structure}\La\ absorption line profiles based on three different atmosphere structures, using the new broadening theory with (blue) and without (green) metals in the calculation of the equation of state, and using the old line broadening including metals in the equation of state (red). Synthetic spectra were computed from these three atmosphere structures without metals to facilitate the comparison of the resulting \La\ profiles. Whereas the photospheric metals do not affect the structure of the atmosphere, the stronger \La\ absorption resulting from the improved line broadening results in a somewhat stronger line blanketing, which will slightly affect the derived effective temperature, if included in a detailed fit. For the purpose of this paper, we adopt \Teff\ and $\log g$ from Table\,\ref{tab:par}.}
\end{figure}

\begin{figure*}
\centerline{\includegraphics[width=0.8\textwidth]{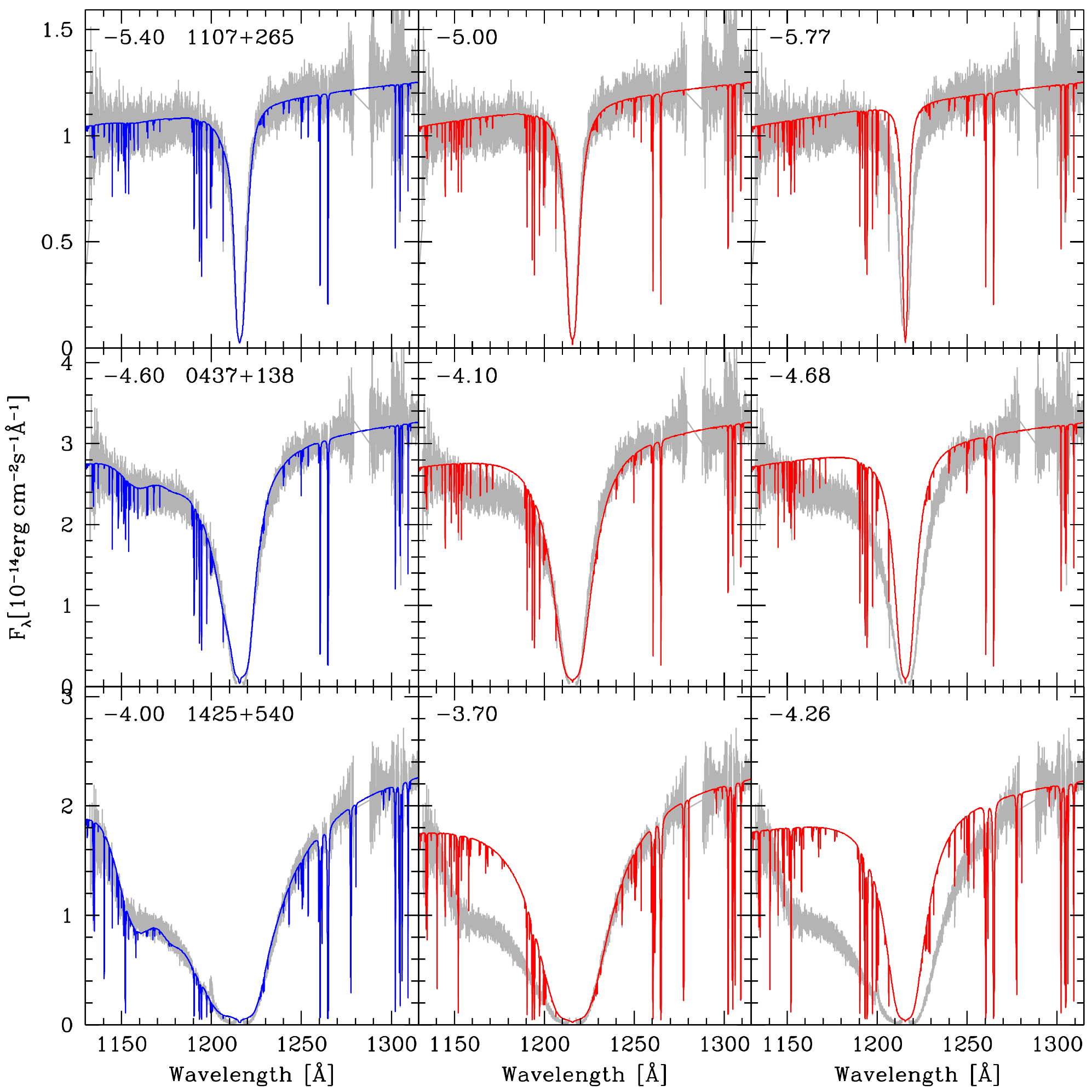}}
\caption{\label{fig:spectra} \textit{HST}/COS spectra of three DB white dwarfs (gray) with the strength of \La\ increasing from top to bottom. The region around \Lines{O}{I}{1302,04} is noisier as we only used the night-side data in this wavelength range. Right panels: model spectra (red) computed with \HHe\ based on a fit to the \Ha\ line, and using \La\ line profiles that include Stark broadening by electrons, H$^+$, and He$^+$ and van der Waals broadening by neutral helium and hydrogen. Middle panels: using the same broadening theory, we varied \HHe\ to reproduce the red wing of \La. The models (red) reproduce well the strength of the observed \La\ line for WD\,1107+265, but clearly fail to fit the observations of WD\,0437+138 and WD\,1425+540. \La. Left panels: fits to the COS spectra using models (blue) including our updated treatment of \La\ broadening by neutral helium result in much better agreement with the observations, and reproduce the depression near 1150\,\AA, a satellite feature of \La. The values of \HHe\ used for the model calculations are given in each panel.}
\end{figure*}

The COS spectra of all 17 stars contain a broad \La\ line, indicating the presence of trace amounts of photospheric hydrogen. The strength of the \La\ line depends on the abundance of hydrogen, \HHe, as well as on the effective temperature \Teff, with the line weakening with increasing \Teff\ for constant values of \HHe. In the case of moderately weak \La\ absorption, the continuum  can be smoothly interpolated across the \La\ line line (WD\,1107+265, top panel in Fig.\,\ref{fig:spectra}). With increasing strength of the \La\ profile, the blue wing becomes increasingly suppressed, and a broad dip centred near 1150\,\AA\ becomes noticeable, with the most extreme example being WD\,1425+540 (bottom panel in Fig.\,\ref{fig:spectra}). With the exception of WD\,0840+262, WD\,1557+192 and WD\,1940+374, the DBs in this sample also display absorption from photospheric metals, which will be discussed in a separate paper (Farihi et al. in prep).

In addition to the \textit{HST} observations, we obtained optical intermediate resolution spectroscopy of WD\,0110--565 and WD\,1349--230 using X-Shooter on the Very Large Telescope, which was used to determine the effective temperatures, surface gravities, as well as \HHe\ from the \Ha\ line.

\begin{table*}
  \centering
\caption{Atmospheric parameters of the 17 DB white dwarfs. Effective temperatures, surface gravities, and hydrogen abundances measured from \Ha\ are taken from Rolland et al. (\citeyear{rollandetal18-1}; R18) or derived from our X-Shooter spectra (XS). Hydrogen abundances were measured from the \La\ absorption line in the COS spectra using \La\ line profiles that include Stark broadening by electrons, H$^+$, and He$^+$ and van der Waals broadening by neutral helium and hydrogen (old) and using our improved unified broadening calculations (new). Stars with photospheric metals are flagged as $z$, and interstellar absorption is likely to contribute to the observed \La\ profile WD\,1349--230, WD\,1557+192, and WD\,2354+159, flagged as $i$.}
\begin{tabular}{lcccccl}
\hline
\hline
WD & \Teff\ & $\log g$ & \multicolumn{3}{c}{\HHe} & Ref.  \\
& [K] & c.g.s. & \Ha & \La\ new & \La\ old & \\
\hline
0100--068$^z$ & $19820\pm~531$ & $8.06\pm0.04$ & $-5.14\pm1.06$ & $-4.80\pm0.15$ & --4.80 & R18\\
0110--565$^z$    & $19124\pm~~16$ & $8.17\pm0.01$ & $-4.20\pm0.20$ & $-4.10\pm0.10$ & --4.10 & XS\\
0125--236$^z$    & $16550\pm~436$ & $8.24\pm0.07$ & $-5.21\pm0.32$ & $-5.00\pm0.15$ & --4.80 & R18\\
0435+410$^z$     & $16790\pm~408$ & $8.18\pm0.08$ & $-4.21\pm0.07$ & $-4.10\pm0.15$ & --4.00 & R18\\
0437+138$^z$     & $15120\pm~361$ & $8.25\pm0.07$ & $-4.68\pm0.06$ & $-4.60\pm0.10$ & --4.10 & R18\\
0840+262     & $17700\pm~863$ & $8.28\pm0.05$ & $-4.18\pm0.06$ & $-4.15\pm0.15$ & --3.95 & R18\\
1107+265$^z$     & $15130\pm~357$ & $8.11\pm0.06$ & $-5.77\pm0.46$ & $-5.40\pm0.15$ & --5.00 & R18\\
1349--230$^{z,i}$& $17905\pm~124$ & $8.05\pm0.01$ & $-4.90\pm0.20$ & $-3.90\pm0.20$ & --3.90 & XS\\
1352+004$^z$     & $13980\pm~340$ & $8.05\pm0.09$ & $-5.31\pm0.17$ & $-5.10\pm0.15$ & --4.70 & R18\\
1425+540$^z$     & $14410\pm~341$ & $7.89\pm0.06$ & $-4.26\pm0.03$ & $-4.00\pm0.20$ & --3.70 & R18\\
1557+192$^i$ & $19510\pm~546$ & $8.15\pm0.05$ & $-4.30\pm0.26$ & $-3.40\pm0.30$ & --3.20 & R18\\
1644+198$^z$     & $15210\pm~360$ & $8.14\pm0.06$ & $-5.68\pm0.39$ & $-5.20\pm0.10$ & --4.90 & R18\\
1822+410$^z$     & $16230\pm~383$ & $8.00\pm0.06$ & $-4.45\pm0.06$ & $-4.40\pm0.20$ & --4.20 & R18\\
1940+374     & $16850\pm~406$ & $8.07\pm0.09$ & $-5.97\pm1.50$ & $-5.60\pm0.15$ & --5.10 & R18\\
2144--079$^z$    & $16340\pm~408$ & $8.18\pm0.05$ & $<-6.22$       & $-6.35\pm0.20$ & --6.20 & R18\\
2229+139$^z$     & $14870\pm~352$ & $8.15\pm0.06$ & $-4.91\pm0.08$ & $-4.70\pm0.15$ & --4.30 & R18\\
2354+159$^{z,i}$ & $24830\pm1670$ & $8.15\pm0.04$ & $<-4.59$       & $-3.60\pm0.20$ & --3.50 & R18\\
\hline
\end{tabular}
\label{tab:par}
\end{table*}

\section{Spectroscopic analysis}

We modelled the \textit{HST}/COS observations using theoretical spectra calculated using the physics and algorithms that are described in \citet{koester10-1}, and which were recently used for the analysis of DB white dwarfs in \citet{koester+kepler15-1}. As the focus of this study is a comparison of the hydrogen abundances measured from \Ha\ and \La, we decided to adopt the atmospheric parameters from \citet{rollandetal18-1} who presented a homogeneous studies of a large sample of DB white dwarfs. Two exceptions to this were WD\,0110--565 and WD\,1349--230, where \Teff,  $\log g$, and \HHe\ based on \Ha\ were determined from our X-Shooter spectroscopy. We compared model spectra computed for these parameters (Table\,\ref{tab:par}) with the PanSTARRS photometry available for all stars except WD\,0110--565, and found good agreement when allowing for small amounts of reddening, $E(B-V)<0.02$. 

Using the values of \HHe\ listed in Table\,\ref{tab:par}, we computed synthetic \La\ profiles using a broadening theory for \La\ which considers Stark broadening by electrons, H$^+$, and He$^+$ ions (\citealt{tremblay+bergeron09-1}, Tremblay priv. comm.), convolved with van der Waals broadening by neutral He and H atoms. These predicted \La\ lines are significantly narrower than the observed ones for all objects (Fig.\,\ref{fig:spectra} and \ref{fig:a1}--\ref{fig:a3}, right panels) except in WD\,0840+262, WD\,1940+374, and WD\,2144--079. An attempt to match the width of the observed \La\ profiles requires hydrogen abundances up to a factor ten larger than those derived from \Ha\ (Fig.\,\ref{fig:spectra} and \ref{fig:a1}--\ref{fig:a3}, middle panels), and still fails to reproduce the shape of the blue wing of \La. This problem was already noted in WD\,1425+540 by \citet{xuetal17-1} and studied in detail by \citet{genest-beaulieu+bergeron17-1}, who suggested that an inhomogeneous distribution of hydrogen in the atmosphere, with \HHe\ increasing in the outer layers, might explain the strong observed \La\ profile. Here we argue that the most likely origin of this discrepancy is that an inadequate broadening theory of \La\ by neutral helium is used in the current model atmosphere codes. 

\begin{figure}
\centerline{\includegraphics[width=\columnwidth]{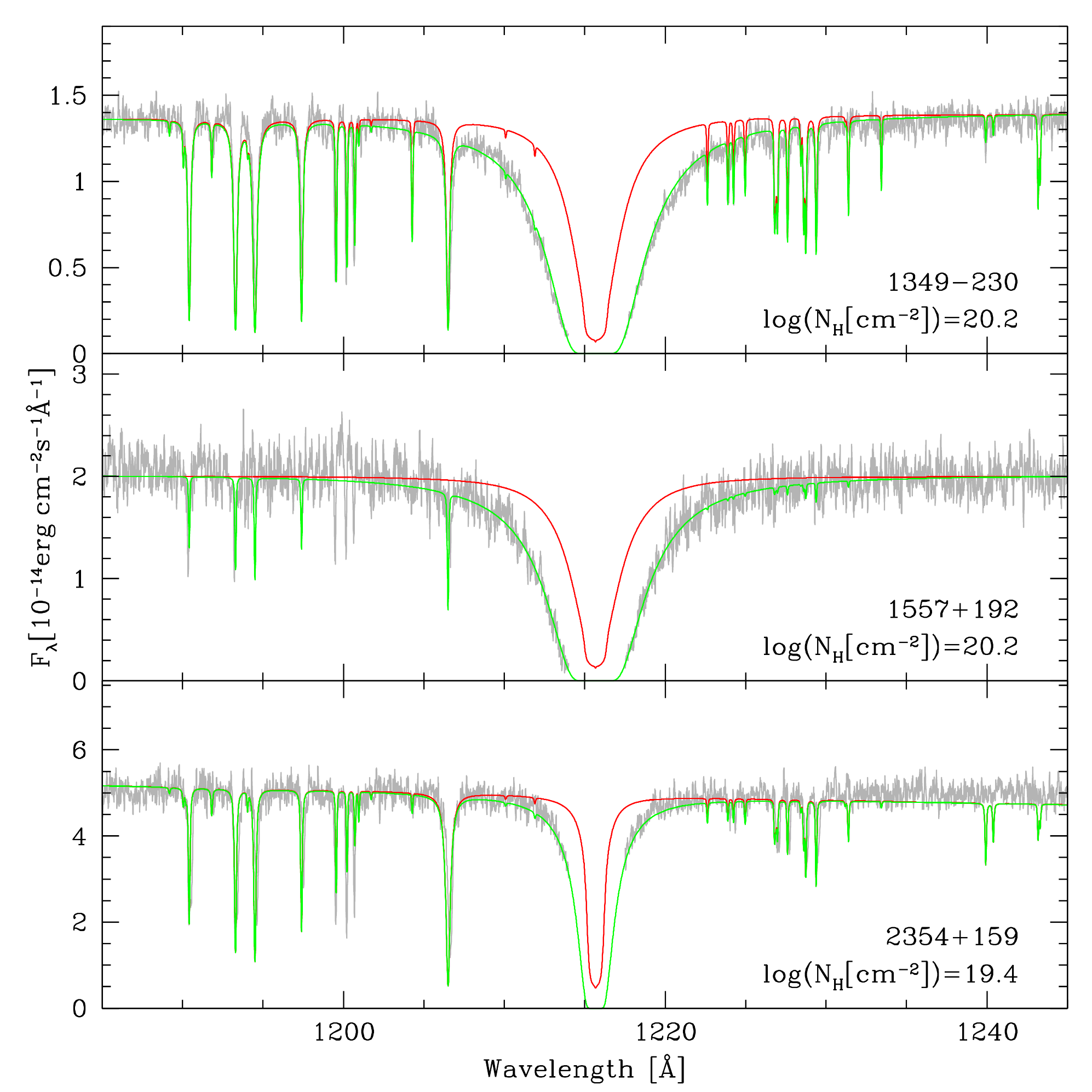}}
\caption{\label{fig:ism} The values of \HHe\ derived from the \La\ absorption in the COS spectra of WD\,1349--230, WD\,1557+192, and WD\,2354+159 are significantly higher than those measured from \Ha. Shown in red are model spectra computed for \Teff\ and $\log g$ from Table\,\ref{tab:par}, and adopting \HHe\ determined from \Ha. These three stars are also among the most distant in our sample, and allowing for a small amount of absorption by interstellar neutral hydrogen column density, quoted in the lower right of each panel, provides a satisfactory fit (green).}
\end{figure}

We had already developed a unified profile calculation \citep{koester+wolff00-1}, but that study was concerned with the analysis of cool ($\Teff<9000$\,K) helium-rich white dwarfs with practically no flux below 1500\,\AA. We therefore used a simplified theory for only one temperature and one perturber density and scaled this profile with the neutral helium density. This treatment is not appropriate to describe the whole \La\ profile, including the blue wing visible in the spectra of the present sample of hotter DB white dwarfs. Improved line profile calculations were performed by \citet{allard+christova09-1}, who predicted a small satellite feature in the blue wing of \La\ near 1150\,\AA, coinciding with the depression detected in the COS spectra of the white dwarfs with the strongest \La\ absorption. Since these authors did not provide numerical tables, we have repeated the profile calculation based on the unified theory by Allard and co-workers \citep{allard+kielkopf82-1, allard+koester92-1, allardetal99-1}, but using our own complete new implementation described in \citet{hollandsetal17-1}. Adiabatic potential curves and dipole moments were obtained from \citet{theodorakopoulosetal87-1, theodorakopoulosetal84-1} and \citet{belyaev15-1}. We calculated tables covering the whole temperature, density, and wavelength ranges where this mechanism is important. Figure\,\ref{fig:profiles} compares these new line profiles to our old treatment of \La\ broadening described above for a range of effective temperatures and hydrogen abundances, and illustrates the increasing importance of broadening by neutral helium both for decreasing \Teff\ and increasing \HHe. Given the strength of this additional broadening of the blue wing of \La, and the fact that most of the DBs also have photospheric metals, we evaluated their influence on the atmospheric structure. Using WD\,1425+540 as test case, we computed the structure of the atmosphere, adopting $\HHe=-4.0$, (1) using the new \La\ line profile and including the best-fit metal abundances in the equation of state, (2) using the old \La\ line profile and including the best-fit metal abundances in the equation of state, and (3) using the new \La\ line profile but no metals. We then computed synthetic spectra without metals, and compared the resulting \La\ absorption lines. Figure.\,\ref{fig:structure} and illustrate that the inclusion of metals has no noticeable effect on the atmospheric structure, but that the stronger \La\ line resulting from the new broadening theory causes a small amount of additional line blanketing. We conclude that the atmospheric parameters of the white dwarfs in our sample can potentially be improved making use of the improved \La\ profiles, though including the \textit{Gaia} Data Release~2 parallaxes \citep{gaiaetal18-1} in the fit may have a stronger effect. However, such a detailed analysis is beyond the scope of this paper, and the relatively small expected changes in \Teff\ and $\log g$ will not affect our main conclusions.

With these new line profiles implemented in our atmosphere code, we fitted the COS spectra of the 17 DB white dwarfs. We adopted, as before, \Teff\ and $\log g$ listed in Table\,\ref{tab:par} and included photospheric metals, where present, in the equation of state computation of the atmospheric structure. We found overall reasonably good agreement between the slope and absolute fluxes of these models, and the COS spectra.\footnote{One exception is WD\,2354+159, where the slope of the model is clearly too blue compared to the COS data (Fig.\,\ref{fig:a3}). We note that this star has the largest uncertainty on the \Teff\ derived from the optical spectroscopy (Table\,\ref{tab:par}). We find evidence for non-negligible amounts of extinction from the strength of the observed \La\ profile (see below), and conclude that the atmospheric parameters WD\,2354+159 remain subject to a careful re-examination taking into account all available spectroscopy, photometry, and astrometry.}. We included absorption by photospheric metals in the calculation of the synthetic spectra where appropriate, and will report the resulting abundances in a separate paper (Farihi et al. in prep). As a result of the new theory, the fits of the asymmetric blue wing and the satellite feature are significantly improved, in particular for the stars with the strongest \La\ lines (Fig.\,\ref{fig:spectra}, Fig.\,\ref{fig:a1}-\ref{fig:a3}, left panels), and we found that the hydrogen abundances measured from the optical and ultraviolet lines are in satisfactory agreement for most objects (Table\,\ref{tab:par}).

Three stars still show significantly higher \HHe\ values measured from \La\ compared to those derived from \Ha: WD\,1349--230, WD\,1557+192, and WD\,2354+159. One possible cause for this discrepancy is an additional contribution of interstellar absorption by neutral hydrogen to the observed \La\ profile. We note that WD\,1349--230 ($d=129$\,pc) and WD\,1557+192 ($d=143$\,pc) are the most distant stars in our sample \citep{gaiaetal18-1}, and with $d>100$\,pc, i.e. outside the local bubble, some interstellar absorption is to be expected. WD\,2354+159 ($d=96$\,pc) is the fourth most distant star in this sample, and also being the hottest one, susceptible to a small contribution from interstellar hydrogen absorption. To test this hypothesis, we re-fitted the COS spectra of these three white dwarfs, allowing for interstellar absorption. We found that the observed \La\ lines of WD\,1349--230 and WD\,1557+192 were well-fitted with $\log(N_\mathrm{H_I}[\mathrm{cm^{-2}}])=20.2$, corresponding to a reddening of $E(B-V)=0.03$ (Fig.\,\ref{fig:ism}). For WD\,2354+159, $\log(N_\mathrm{H_I}[\mathrm{cm^{-2}}])$, corresponding to $E(B-V)=0.005$ was sufficient to match the width of the \La\ profile in the COS spectrum. These small amounts of reddening are consistent with the extinction towards this star estimated from the 3D reddening maps of \citet{lallementetal14-1} and \citet{capitanioetal17-1}.

\section{Conclusions}
We have developed new line profile calculations for \La\ which account for broadening by neutral helium. When fitting \textit{HST} COS spectroscopy of 17 DB white dwarfs using these new line profiles, we find good agreement between \HHe\ measured from \La\ and \Ha, resolving discrepancies that were present in analyses that did not properly account for the additional line broadening. Three stars still show a larger value of \HHe\ measured from \La\ compared to that derived from \Ha, and we argue that a small contribution of interstellar absorption from neutral hydrogen is a plausible cause for this discrepancy. 

\section*{Acknowledgements}
The research leading to these results has received funding from the European Research Council under the European Union's Seventh Framework Programme (FP/2007- 2013) / ERC Grant Agreement n. 320964 (WDTracer). OT was also supported by a Leverhulme Trust Research Project Grant. This work is based on observations made with the NASA/ESA Hubble Space Telescope, obtained at the Space Telescope Science Institute, which is operated by the Association of Universities for Research in Astronomy, Inc., under NASA contract NAS 5-26555. These observations are associated with programmes \#12474, \#13453 and \#14597. Based on observations made with ESO Telescopes at the La Silla Paranal Observatory under programme ID 087.D-0858. We thank Pierre Bergeron for a constructive referee report.




\bibliographystyle{mnras}
\bibliography{aamnem99,a-k,l-z,proceedings} 

\begin{thebibliography}{}
\makeatletter
\relax
\def\mn@urlcharsother{\let\do\@makeother \do\$\do\&\do\#\do\^\do\_\do\%\do\~}
\def\mn@doi{\begingroup\mn@urlcharsother \@ifnextchar [ {\mn@doi@}
  {\mn@doi@[]}}
\def\mn@doi@[#1]#2{\def\@tempa{#1}\ifx\@tempa\@empty \href
  {http://dx.doi.org/#2} {doi:#2}\else \href {http://dx.doi.org/#2} {#1}\fi
  \endgroup}
\def\mn@eprint#1#2{\mn@eprint@#1:#2::\@nil}
\def\mn@eprint@arXiv#1{\href {http://arxiv.org/abs/#1} {{\tt arXiv:#1}}}
\def\mn@eprint@dblp#1{\href {http://dblp.uni-trier.de/rec/bibtex/#1.xml}
  {dblp:#1}}
\def\mn@eprint@#1:#2:#3:#4\@nil{\def\@tempa {#1}\def\@tempb {#2}\def\@tempc
  {#3}\ifx \@tempc \@empty \let \@tempc \@tempb \let \@tempb \@tempa \fi \ifx
  \@tempb \@empty \def\@tempb {arXiv}\fi \@ifundefined
  {mn@eprint@\@tempb}{\@tempb:\@tempc}{\expandafter \expandafter \csname
  mn@eprint@\@tempb\endcsname \expandafter{\@tempc}}}

\bibitem[\protect\citeauthoryear{{Allard} \& {Christova}}{{Allard} \&
  {Christova}}{2009}]{allard+christova09-1}
{Allard} N.~F.,  {Christova} M.,  2009, \mn@doi [New Astronomy Reviews]
  {10.1016/j.newar.2009.07.007}, \href {2009NewAR..53..252A} {53, 252}

\bibitem[\protect\citeauthoryear{{Allard} \& {Kielkopf}}{{Allard} \&
  {Kielkopf}}{1982}]{allard+kielkopf82-1}
{Allard} N.,  {Kielkopf} J.,  1982, \mn@doi [Reviews of Modern Physics]
  {10.1103/RevModPhys.54.1103}, \href {1982RvMP...54.1103A} {54, 1103}

\bibitem[\protect\citeauthoryear{{Allard} \& {Koester}}{{Allard} \&
  {Koester}}{1992}]{allard+koester92-1}
{Allard} N.~F.,  {Koester} D.,  1992, A\&A, \href {1992A&A...258..464A} {258,
  464}

\bibitem[\protect\citeauthoryear{{Allard}, {Royer}, {Kielkopf}  \&
  {Feautrier}}{{Allard} et~al.}{1999}]{allardetal99-1}
{Allard} N.~F.,  {Royer} A.,  {Kielkopf} J.~F.,   {Feautrier} N.,  1999,
  \mn@doi [Phys. Rev. A] {10.1103/PhysRevA.60.1021}, \href
  {1999PhRvA..60.1021A} {60, 1021}

\bibitem[\protect\citeauthoryear{{Althaus}, {Camisassa}, {Miller Bertolami},
  {C{\'o}rsico}  \& {Garc{\'{\i}}a-Berro}}{{Althaus}
  et~al.}{2015}]{althausetal15-1}
{Althaus} L.~G.,  {Camisassa} M.~E.,  {Miller Bertolami} M.~M.,  {C{\'o}rsico}
  A.~H.,   {Garc{\'{\i}}a-Berro} E.,  2015, \mn@doi [A\&A]
  {10.1051/0004-6361/201424922}, \href {2015A&A...576A...9A} {576, A9}

\bibitem[\protect\citeauthoryear{{Beauchamp}, {Wesemael}, {Bergeron}, {Liebert}
   \& {Saffer}}{{Beauchamp} et~al.}{1996}]{beauchampetal96-1}
{Beauchamp} A.,  {Wesemael} F.,  {Bergeron} P.,  {Liebert} J.,   {Saffer}
  R.~A.,  1996, in {Jeffery} C.~S.,  {Heber} U.,  eds,  Astronomical Society of
  the Pacific Conference Series Vol. 96, Hydrogen Deficient Stars. p.~295

\bibitem[\protect\citeauthoryear{{Belyaev}}{{Belyaev}}{2015}]{belyaev15-1}
{Belyaev} A.~K.,  2015, \mn@doi [Phys. Rev. A] {10.1103/PhysRevA.91.062709},
  \href {2015PhRvA..91f2709B} {91, 062709}

\bibitem[\protect\citeauthoryear{{Bergeron} \& {Liebert}}{{Bergeron} \&
  {Liebert}}{2002}]{bergeron+liebert02-1}
{Bergeron} P.,  {Liebert} J.,  2002, \mn@doi [ApJ] {10.1086/338279}, \href
  {2002ApJ...566.1091B} {566, 1091}

\bibitem[\protect\citeauthoryear{{Bergeron}, {Ruiz}  \& {Leggett}}{{Bergeron}
  et~al.}{1997}]{bergeronetal97-1}
{Bergeron} P.,  {Ruiz} M.~T.,   {Leggett} S.~K.,  1997, \mn@doi [ApJS]
  {10.1086/312955}, \href {1997ApJS..108..339B} {108, 339}

\bibitem[\protect\citeauthoryear{{Bergeron} et~al.,}{{Bergeron}
  et~al.}{2011}]{bergeronetal11-1}
{Bergeron} P.,  et~al., 2011, \mn@doi [ApJ] {10.1088/0004-637X/737/1/28}, \href
  {2011ApJ...737...28B} {737, 28}

\bibitem[\protect\citeauthoryear{{Capitanio}, {Lallement}, {Vergely},
  {Elyajouri}  \& {Monreal-Ibero}}{{Capitanio}
  et~al.}{2017}]{capitanioetal17-1}
{Capitanio} L.,  {Lallement} R.,  {Vergely} J.~L.,  {Elyajouri} M.,
  {Monreal-Ibero} A.,  2017, \mn@doi [A\&A] {10.1051/0004-6361/201730831},
  \href {2017A&A...606A..65C} {606, A65}

\bibitem[\protect\citeauthoryear{{Chen} \& {Hansen}}{{Chen} \&
  {Hansen}}{2012}]{chenetal12-1}
{Chen} E.~Y.,  {Hansen} B.~M.~S.,  2012, \mn@doi [ApJ Lett.]
  {10.1088/2041-8205/753/1/L16}, \href {2012ApJ...753L..16C} {753, L16}

\bibitem[\protect\citeauthoryear{{Doherty}, {Gil-Pons}, {Siess}, {Lattanzio}
  \& {Lau}}{{Doherty} et~al.}{2015}]{dohertyetal15-1}
{Doherty} C.~L.,  {Gil-Pons} P.,  {Siess} L.,  {Lattanzio} J.~C.,   {Lau}
  H.~H.~B.,  2015, \mn@doi [MNRAS] {10.1093/mnras/stu2180}, \href
  {2015MNRAS.446.2599D} {446, 2599}

\bibitem[\protect\citeauthoryear{{Farihi}, {Barstow}, {Redfield}, {Dufour}  \&
  {Hambly}}{{Farihi} et~al.}{2010}]{farihietal10-2}
{Farihi} J.,  {Barstow} M.~A.,  {Redfield} S.,  {Dufour} P.,   {Hambly} N.~C.,
  2010, \mn@doi [MNRAS] {10.1111/j.1365-2966.2010.16426.x}, \href
  {2010MNRAS.404.2123F} {404, 2123}

\bibitem[\protect\citeauthoryear{{Farihi}, {G{\"a}nsicke}  \&
  {Koester}}{{Farihi} et~al.}{2013}]{farihietal13-2}
{Farihi} J.,  {G{\"a}nsicke} B.~T.,   {Koester} D.,  2013, \mn@doi [Science]
  {10.1126/science.1239447}, \href {2013Sci...342..218F} {342, 218}

\bibitem[\protect\citeauthoryear{{Gaia Collaboration} et~al.,}{{Gaia
  Collaboration} et~al.}{2018}]{gaiaetal18-1}
{Gaia Collaboration} et~al., 2018, \mn@doi [A\&A]
  {10.1051/0004-6361/201833051}, \href
  {http://adsabs.harvard.edu/abs/2018A%26A...616A...1G} {616, A1}

\bibitem[\protect\citeauthoryear{{G{\"a}nsicke}, {Koester}, {Farihi}, {Girven},
  {Parsons}  \& {Breedt}}{{G{\"a}nsicke} et~al.}{2012}]{gaensickeetal12-1}
{G{\"a}nsicke} B.~T.,  {Koester} D.,  {Farihi} J.,  {Girven} J.,  {Parsons}
  S.~G.,   {Breedt} E.,  2012, \mn@doi [MNRAS]
  {10.1111/j.1365-2966.2012.21201.x}, \href {2012MNRAS.424..333G} {424, 333}

\bibitem[\protect\citeauthoryear{{Garcia-Berro}, {Ritossa}  \&
  {Iben}}{{Garcia-Berro} et~al.}{1997}]{garcia-berroetal97-1}
{Garcia-Berro} E.,  {Ritossa} C.,   {Iben} I.~J.,  1997, \mn@doi [ApJ]
  {10.1086/304444}, \href {1997ApJ...485..765G} {485, 765}

\bibitem[\protect\citeauthoryear{{Genest-Beaulieu} \&
  {Bergeron}}{{Genest-Beaulieu} \&
  {Bergeron}}{2017}]{genest-beaulieu+bergeron17-1}
{Genest-Beaulieu} C.,  {Bergeron} P.,  2017, in {Tremblay} P.-E.,  {Gaensicke}
  B.,   {Marsh} T.,  eds,  Astronomical Society of the Pacific Conference
  Series Vol. 509, 20th European White Dwarf Workshop. p.~201 (\mn@eprint {}
  {1610.08828})

\bibitem[\protect\citeauthoryear{{Gentile Fusillo}, {G{\"a}nsicke}, {Farihi},
  {Koester}, {Schreiber}  \& {Pala}}{{Gentile Fusillo}
  et~al.}{2017}]{gentile-fusilloetal17-1}
{Gentile Fusillo} N.~P.,  {G{\"a}nsicke} B.~T.,  {Farihi} J.,  {Koester} D.,
  {Schreiber} M.~R.,   {Pala} A.~F.,  2017, \mn@doi [MNRAS]
  {10.1093/mnras/stx468}, \href {2017MNRAS.468..971G} {468, 971}

\bibitem[\protect\citeauthoryear{{Giammichele}, {Bergeron}  \&
  {Dufour}}{{Giammichele} et~al.}{2012}]{giammicheleetal12-1}
{Giammichele} N.,  {Bergeron} P.,   {Dufour} P.,  2012, \mn@doi [ApJS]
  {10.1088/0067-0049/199/2/29}, \href {2012ApJS..199...29G} {199, 29}

\bibitem[\protect\citeauthoryear{{Gianninas}, {Dufour}  \&
  {Bergeron}}{{Gianninas} et~al.}{2004}]{gianninasetal04-1}
{Gianninas} A.,  {Dufour} P.,   {Bergeron} P.,  2004, \mn@doi [ApJ Lett.]
  {10.1086/427080}, \href {2004ApJ...617L..57G} {617, L57}

\bibitem[\protect\citeauthoryear{{Hollands}, {Koester}, {Alekseev}, {Herbert}
  \& {G{\"a}nsicke}}{{Hollands} et~al.}{2017}]{hollandsetal17-1}
{Hollands} M.~A.,  {Koester} D.,  {Alekseev} V.,  {Herbert} E.~L.,
  {G{\"a}nsicke} B.~T.,  2017, \mn@doi [MNRAS] {10.1093/mnras/stx250}, \href
  {2017MNRAS.467.4970H} {467, 4970}

\bibitem[\protect\citeauthoryear{{Iben}, {Kaler}, {Truran}  \&
  {Renzini}}{{Iben} et~al.}{1983}]{ibenetal83-1}
{Iben} Jr. I.,  {Kaler} J.~B.,  {Truran} J.~W.,   {Renzini} A.,  1983, \mn@doi
  [ApJ] {10.1086/160631}, \href {1983ApJ...264..605I} {264, 605}

\bibitem[\protect\citeauthoryear{{Jura} \& {Xu}}{{Jura} \&
  {Xu}}{2012}]{jura+xu12-1}
{Jura} M.,  {Xu} S.,  2012, \mn@doi [AJ] {10.1088/0004-6256/143/1/6}, \href
  {2012AJ....143....6J} {143, 6}

\bibitem[\protect\citeauthoryear{{Kawka} \& {Vennes}}{{Kawka} \&
  {Vennes}}{2005}]{kawka+vennes05-1}
{Kawka} A.,  {Vennes} S.,  2005, in {Koester} D.,  {Moehler} S.,  eds,
  Astronomical Society of the Pacific Conference Series Vol. 334, 14th European
  Workshop on White Dwarfs. ASP Conf. Ser. 336, pp 101--106

\bibitem[\protect\citeauthoryear{{Koester}}{{Koester}}{2010}]{koester10-1}
{Koester} D.,  2010, Memorie della Societa Astronomica Italiana, \href
  {2010MmSAI..81..921K} {81, 921}

\bibitem[\protect\citeauthoryear{{Koester} \& {Kepler}}{{Koester} \&
  {Kepler}}{2015}]{koester+kepler15-1}
{Koester} D.,  {Kepler} S.~O.,  2015, \mn@doi [A\&A]
  {10.1051/0004-6361/201527169}, \href {2015A&A...583A..86K} {583, A86}

\bibitem[\protect\citeauthoryear{{Koester} \& {Wolff}}{{Koester} \&
  {Wolff}}{2000}]{koester+wolff00-1}
{Koester} D.,  {Wolff} B.,  2000, A\&A, \href {2000A&A...357..587K} {357, 587}

\bibitem[\protect\citeauthoryear{{Koester}, {Napiwotzki}, {Voss}, {Homeier}  \&
  {Reimers}}{{Koester} et~al.}{2005}]{koesteretal05-1}
{Koester} D.,  {Napiwotzki} R.,  {Voss} B.,  {Homeier} D.,   {Reimers} D.,
  2005, \mn@doi [A\&A] {10.1051/0004-6361:20053058}, \href
  {2005A&A...439..317K} {439, 317}

\bibitem[\protect\citeauthoryear{{Lallement}, {Vergely}, {Valette},
  {Puspitarini}, {Eyer}  \& {Casagrande}}{{Lallement}
  et~al.}{2014}]{lallementetal14-1}
{Lallement} R.,  {Vergely} J.-L.,  {Valette} B.,  {Puspitarini} L.,  {Eyer} L.,
    {Casagrande} L.,  2014, \mn@doi [A\&A] {10.1051/0004-6361/201322032}, \href
  {2014A&A...561A..91L} {561, A91}

\bibitem[\protect\citeauthoryear{{Liebert}, {Gresham}, {Hege}  \&
  {Strittmatter}}{{Liebert} et~al.}{1979}]{liebertetal79-1}
{Liebert} J.,  {Gresham} M.,  {Hege} E.~K.,   {Strittmatter} P.~A.,  1979,
  \mn@doi [AJ] {10.1086/112584}, \href {1979AJ.....84.1612L} {84, 1612}

\bibitem[\protect\citeauthoryear{{Liebert}, {Wesemael}, {Sion}  \&
  {Wegner}}{{Liebert} et~al.}{1984}]{liebertetal84-1}
{Liebert} J.,  {Wesemael} F.,  {Sion} E.~M.,   {Wegner} G.,  1984, \mn@doi
  [ApJ] {10.1086/161740}, \href {1984ApJ...277..692L} {277, 692}

\bibitem[\protect\citeauthoryear{{Manseau}, {Bergeron}  \& {Green}}{{Manseau}
  et~al.}{2016}]{manseauetal16-1}
{Manseau} P.~M.,  {Bergeron} P.,   {Green} E.~M.,  2016, \mn@doi [ApJ]
  {10.3847/1538-4357/833/2/127}, \href {2016ApJ...833..127M} {833, 127}

\bibitem[\protect\citeauthoryear{{Raddi}, {G{\"a}nsicke}, {Koester}, {Farihi},
  {Hermes}, {Scaringi}, {Breedt}  \& {Girven}}{{Raddi}
  et~al.}{2015}]{raddietal15-1}
{Raddi} R.,  {G{\"a}nsicke} B.~T.,  {Koester} D.,  {Farihi} J.,  {Hermes}
  J.~J.,  {Scaringi} S.,  {Breedt} E.,   {Girven} J.,  2015, \mn@doi [MNRAS]
  {10.1093/mnras/stv701}, \href {2015MNRAS.450.2083R} {450, 2083}

\bibitem[\protect\citeauthoryear{{Rolland}, {Bergeron}  \&
  {Fontaine}}{{Rolland} et~al.}{2018}]{rollandetal18-1}
{Rolland} B.,  {Bergeron} P.,   {Fontaine} G.,  2018, \mn@doi [ApJ]
  {10.3847/1538-4357/aab713}, \href {2018ApJ...857...56R} {857, 56}

\bibitem[\protect\citeauthoryear{{Schatzman}}{{Schatzman}}{1948}]{schatzman48-1}
{Schatzman} E.,  1948, \mn@doi [Nat] {10.1038/161061b0}, \href
  {1948Natur.161R..61S} {161, 61}

\bibitem[\protect\citeauthoryear{{Sion}, {Szkody}, {Cheng}, {G{\"a}nsicke}  \&
  {Howell}}{{Sion} et~al.}{2003}]{sionetal03-1}
{Sion} E.~M.,  {Szkody} P.,  {Cheng} F.,  {G{\"a}nsicke} B.~T.,   {Howell}
  S.~B.,  2003, ApJ, \href {2003ApJ...583..907S} {583, 907}

\bibitem[\protect\citeauthoryear{{Smartt}, {Eldridge}, {Crockett}  \&
  {Maund}}{{Smartt} et~al.}{2009}]{smarttetal09-1}
{Smartt} S.~J.,  {Eldridge} J.~J.,  {Crockett} R.~M.,   {Maund} J.~R.,  2009,
  \mn@doi [MNRAS] {10.1111/j.1365-2966.2009.14506.x}, \href
  {2009MNRAS.395.1409S} {395, 1409}

\bibitem[\protect\citeauthoryear{{Theodorakopoulos}, {Farantos}, {Buenker}  \&
  {Peyerimhoff}}{{Theodorakopoulos} et~al.}{1984}]{theodorakopoulosetal84-1}
{Theodorakopoulos} G.,  {Farantos} S.~C.,  {Buenker} R.~J.,   {Peyerimhoff}
  S.~D.,  1984, \mn@doi [Journal of Physics B Atomic Molecular Physics]
  {10.1088/0022-3700/17/8/008}, \href {1984JPhB...17.1453T} {17, 1453}

\bibitem[\protect\citeauthoryear{{Theodorakopoulos}, {Petsalakis}, {Nicolaides}
   \& {Buenker}}{{Theodorakopoulos} et~al.}{1987}]{theodorakopoulosetal87-1}
{Theodorakopoulos} G.,  {Petsalakis} I.~D.,  {Nicolaides} C.~A.,   {Buenker}
  R.~J.,  1987, \mn@doi [Journal of Physics B Atomic Molecular Physics]
  {10.1088/0022-3700/20/11/006}, \href {1987JPhB...20.2339T} {20, 2339}

\bibitem[\protect\citeauthoryear{{Tremblay} \& {Bergeron}}{{Tremblay} \&
  {Bergeron}}{2009}]{tremblay+bergeron09-1}
{Tremblay} P.-E.,  {Bergeron} P.,  2009, \mn@doi [ApJ]
  {10.1088/0004-637X/696/2/1755}, \href {2009ApJ...696.1755T} {696, 1755}

\bibitem[\protect\citeauthoryear{{Voss}, {Koester}, {Napiwotzki}, {Christlieb}
  \& {Reimers}}{{Voss} et~al.}{2007}]{vossetal07-1}
{Voss} B.,  {Koester} D.,  {Napiwotzki} R.,  {Christlieb} N.,   {Reimers} D.,
  2007, \mn@doi [A\&A] {10.1051/0004-6361:20077285}, \href
  {2007A&A...470.1079V} {470, 1079}

\bibitem[\protect\citeauthoryear{{Xu}, {Zuckerman}, {Dufour}, {Young}, {Klein}
  \& {Jura}}{{Xu} et~al.}{2017}]{xuetal17-1}
{Xu} S.,  {Zuckerman} B.,  {Dufour} P.,  {Young} E.~D.,  {Klein} B.,   {Jura}
  M.,  2017, \mn@doi [ApJ Lett.] {10.3847/2041-8213/836/1/L7}, \href
  {2017ApJ...836L...7X} {836, L7}

\makeatother
\end{thebibliography}

\appendix

\section{Extra plots}

\begin{figure*}
\centerline{\includegraphics[width=0.9\textwidth]{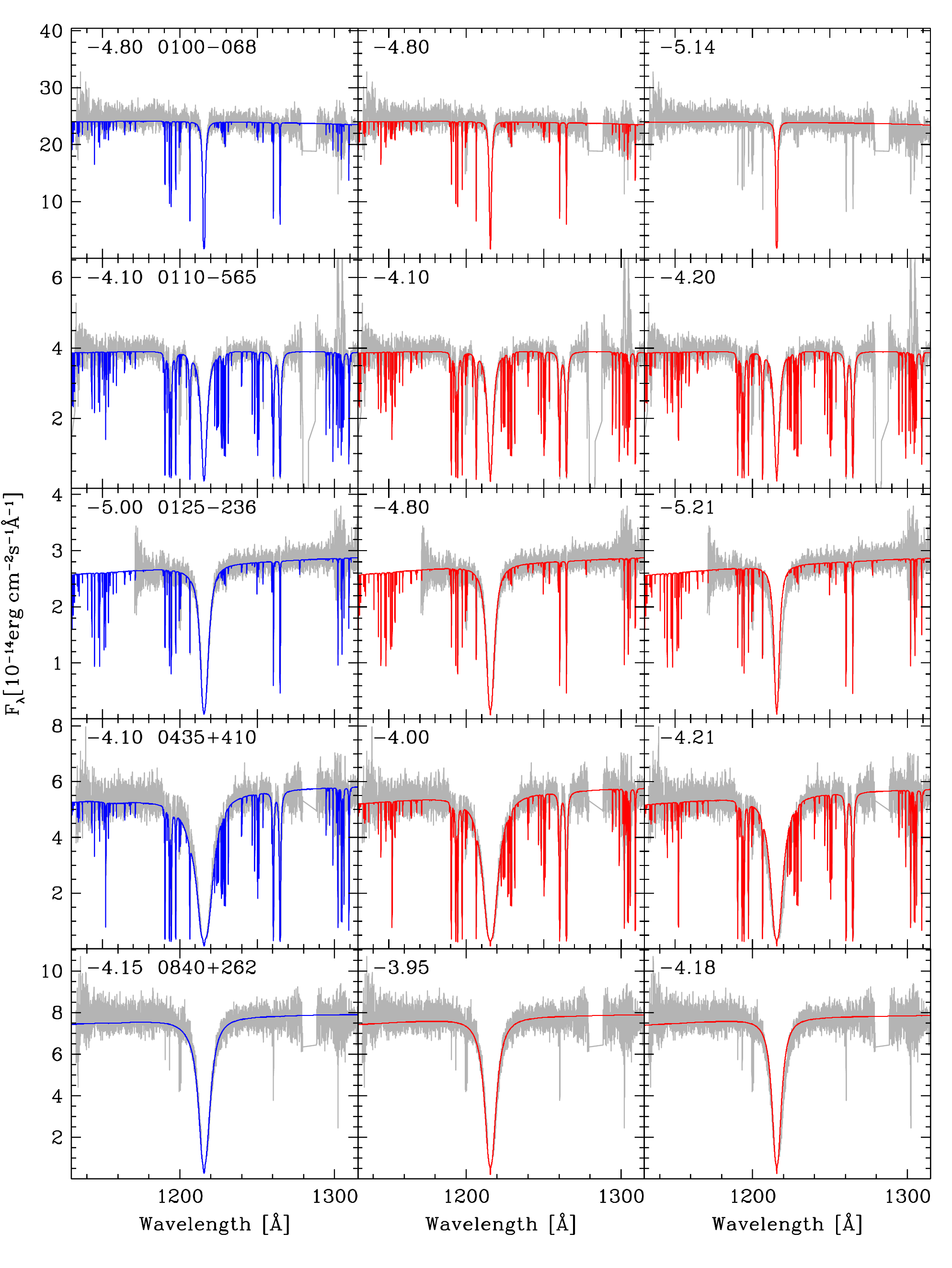}}
\caption{\label{fig:a1} Same as Fig.\,\ref{fig:spectra} for the full DB sample analysed in this paper.}
\end{figure*}

\begin{figure*}
\centerline{\includegraphics[width=0.9\textwidth]{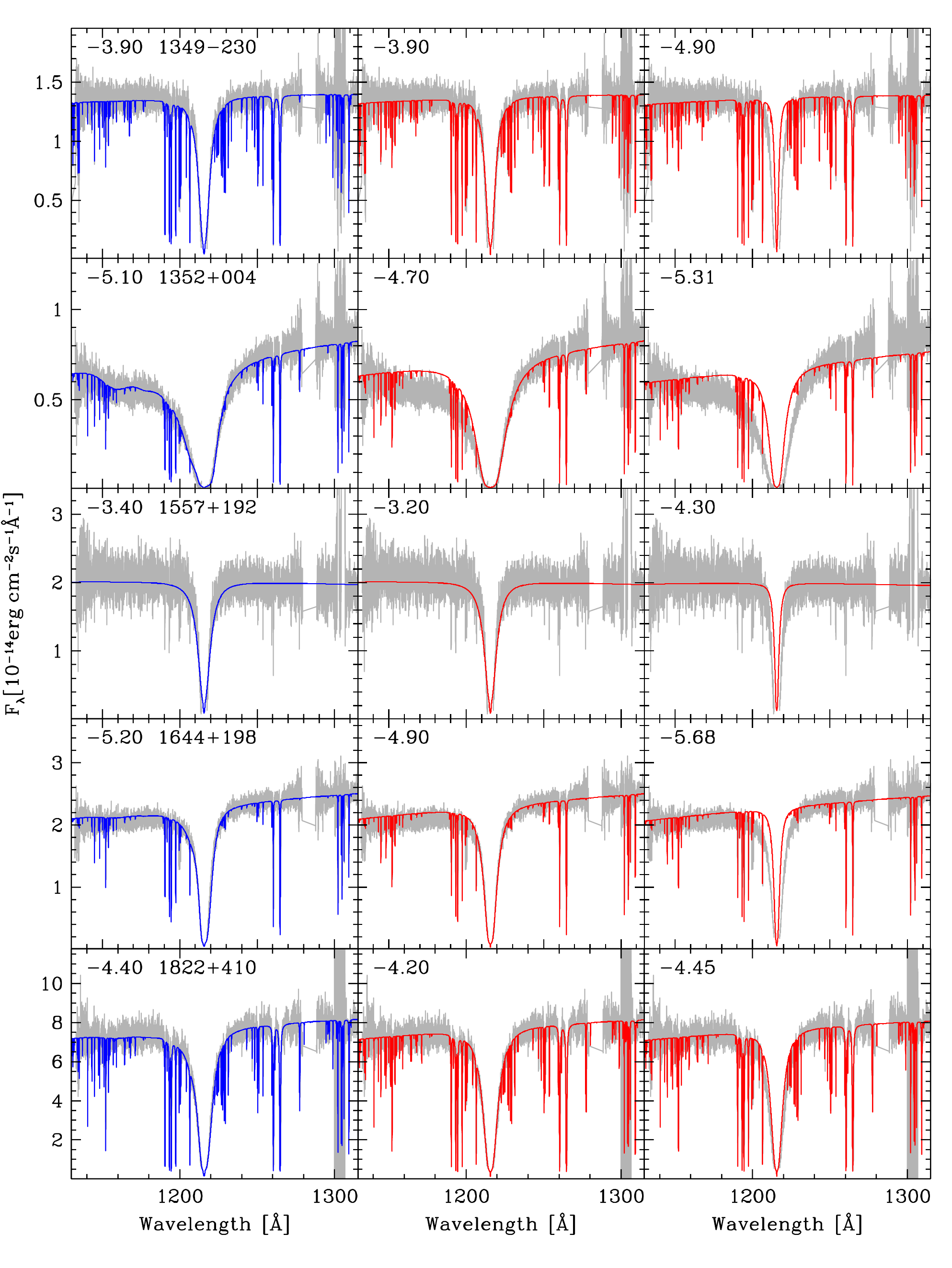}}
\caption{\label{fig:a2} Same as Fig.\,\ref{fig:spectra} for the full DB sample analysed in this paper.}
\end{figure*}

\begin{figure*}
\centerline{\includegraphics[width=0.9\textwidth]{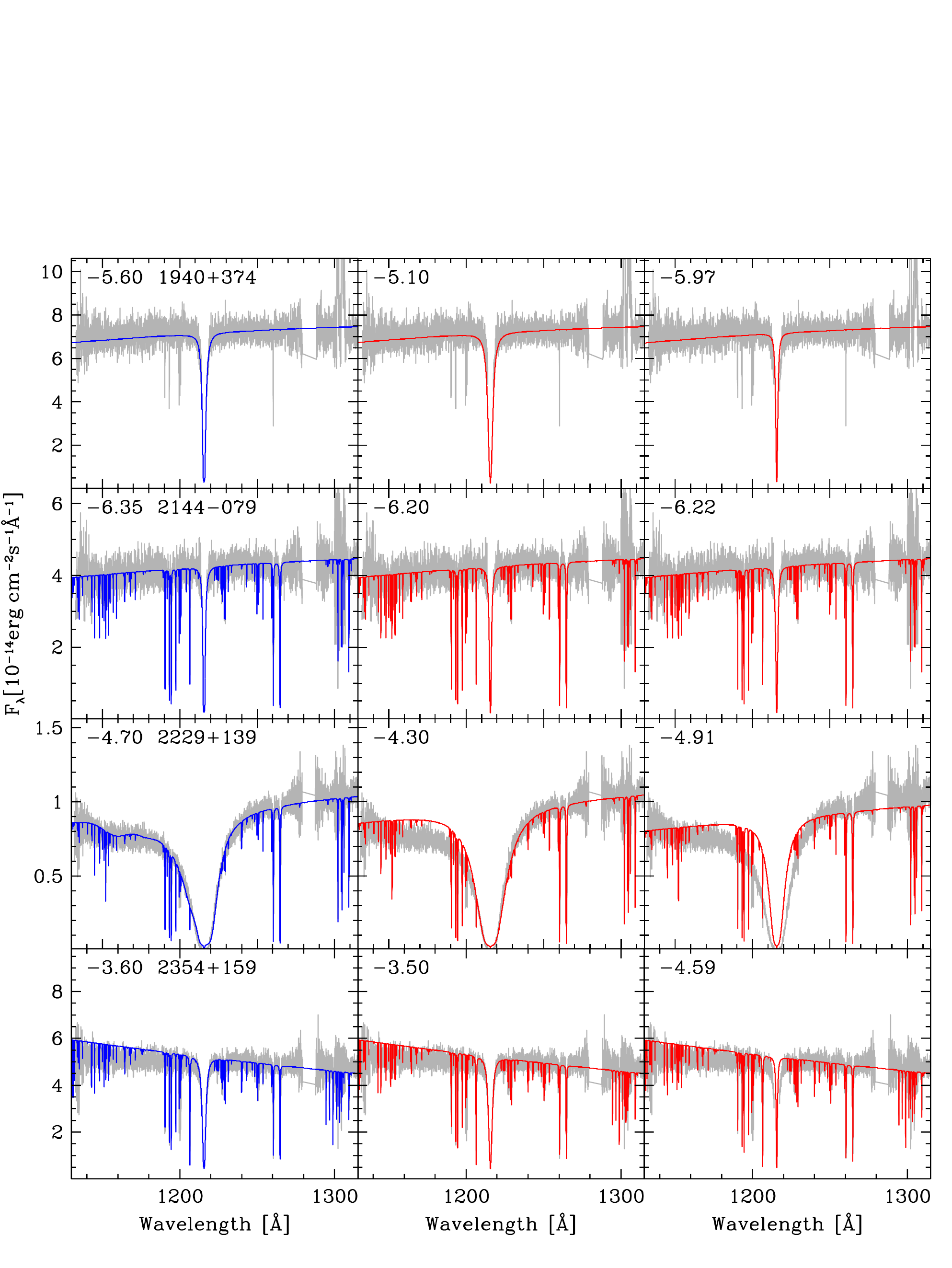}}
\caption{\label{fig:a3} Same as Fig.\,\ref{fig:spectra} for the full DB sample analysed in this paper.}
\end{figure*}

\bsp	
\label{lastpage}
\end{document}